\documentclass[sigconf]{acmart}

\AtBeginDocument{%
  \providecommand\BibTeX{{%
    \normalfont B\kern-0.5em{\scshape i\kern-0.25em b}\kern-0.8em\TeX}}}

\usepackage{multirow}
\usepackage{subfigure}
\usepackage{graphicx}
\usepackage{makecell}

\begin{document}

\title{Open-vocabulary Auditory Neural Decoding Using fMRI-prompted LLM}


\author{Xiaoyu Chen}
\authornote{Both authors contributed equally to this research.}
\orcid{0000-0001-8945-3150}
\affiliation{%
	\institution{Laboratory of Brain Atlas and Brain-Inspired Intelligence, State Key Laboratory of Multimodal Artificial Intelligence Systems, CASIA}
	\institution{School of Artificial Intelligence, University of Chinese Academy of Science}
	\city{Beijing}
	\country{China}
}
\email{chenxiaoyu2022@ia.ac.cn}

\author{Changde Du}
\authornotemark[1]
\orcid{0000-0002-0084-433X}
\affiliation{%
	\institution{Laboratory of Brain Atlas and Brain-Inspired Intelligence, State Key Laboratory of Multimodal Artificial Intelligence Systems, CASIA}
	\city{Beijing}
	\country{China}
}
\email{changde.du@ia.ac.cn}

\author{Liu Che}
\affiliation{%
	\institution{Laboratory of Brain Atlas and Brain-Inspired Intelligence, State Key Laboratory of Multimodal Artificial Intelligence Systems, CASIA}
	\institution{University of Chinese Academy of Science}
	\city{Beijing}
	\country{China}
}
\email{liuche2022@ia.ac.cn}

\author{Yizhe Wang}
\affiliation{%
	\institution{CASIA}
	\city{Beijing}
	\country{China}
}
\email{yizhe.wang@ia.ac.cn}

\author{Huiguang He}
\orcid{0000-0002-0684-1711}
\authornote{corresponding author.}
\affiliation{%
	\institution{Laboratory of Brain Atlas and Brain-Inspired Intelligence, State Key Laboratory of Multimodal Artificial Intelligence Systems, CASIA}
	\institution{School of Artificial Intelligence, University of Chinese Academy of Science}
	\city{Beijing}
	\country{China}
}
\email{huiguang.he@ia.ac.cn}


\begin{abstract}
  Decoding language information from brain signals represents a vital research area within brain-computer interfaces, particularly in the context of deciphering the semantic information from the fMRI signal. However, many existing efforts concentrate on decoding small vocabulary sets, leaving space for the exploration of open vocabulary continuous text decoding. In this paper, we introduce a novel method, the \textbf{Brain Prompt GPT (BP-GPT)}. By using the brain representation that is extracted from the fMRI as a prompt, our method can utilize GPT-2 to decode fMRI signals into stimulus text. Further, we introduce a text-to-text baseline and align the fMRI prompt to the text prompt. By introducing the text-to-text baseline, our BP-GPT can extract a more robust brain prompt and promote the decoding of pre-trained LLM. We evaluate our BP-GPT on the open-source auditory semantic decoding dataset and achieve a significant improvement up to $4.61\%$ on METEOR and $2.43\%$ on BERTScore across all the subjects compared to the state-of-the-art method. The experimental results demonstrate that using brain representation as a prompt to further drive LLM for auditory neural decoding is feasible and effective.
\end{abstract}


\keywords{Do, Not, Us, This, Code, Put, the, Correct, Terms, for,
  Your, Paper}



\maketitle

\section{Introduction}
“The limits of my language mean the limits of my world” - Ludwig Wittgenstein. Wittgenstein’s statement refers to the standpoint that a person's entire understanding of the world is reflected in the things they can describe in language. So it is important for human-centric artificial intelligence, for example, the brain-computer interface, to understand the language information from the human brain. 

\begin{figure}[t]
    \centering
    \includegraphics[width=\linewidth]{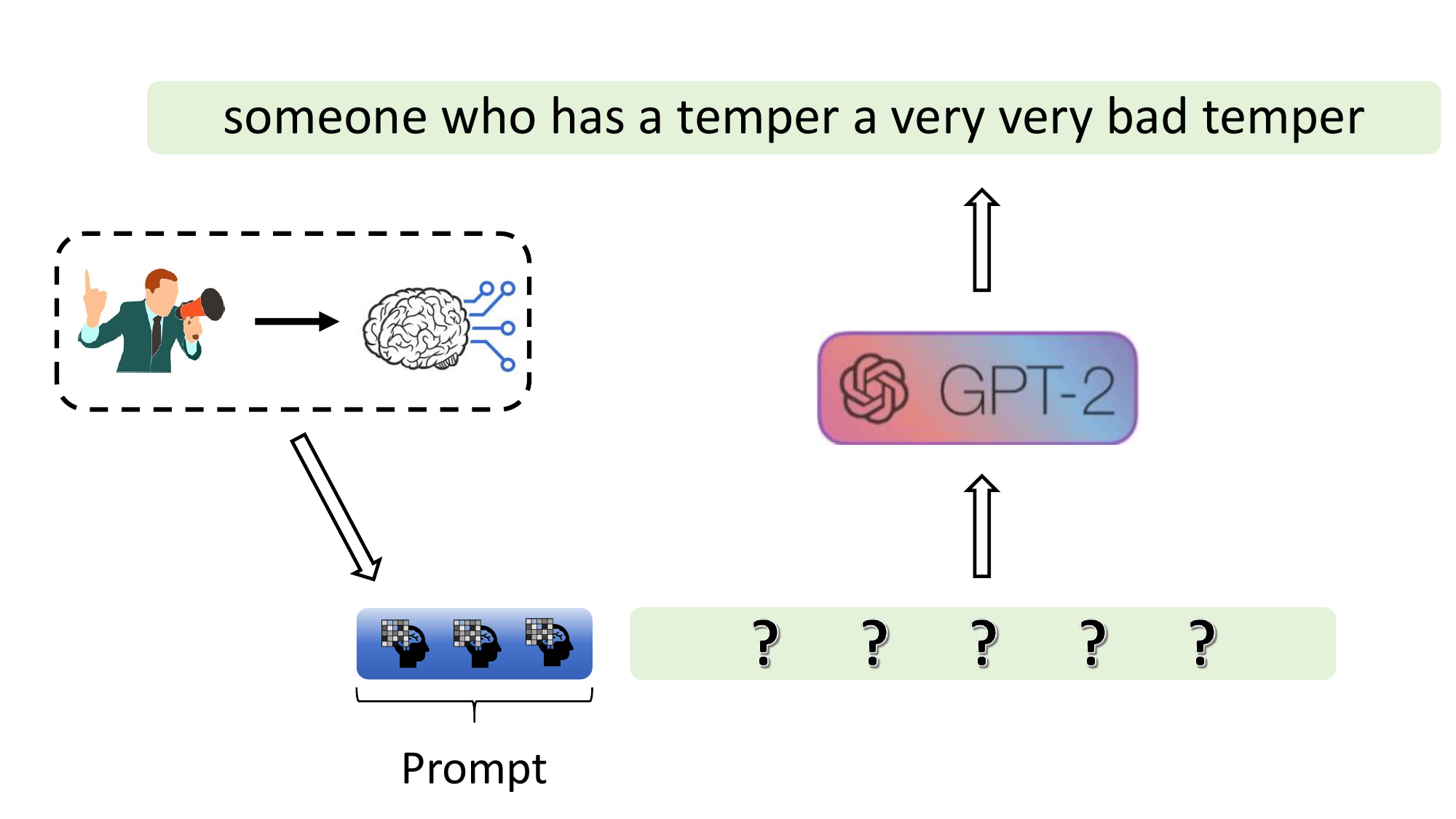}
    \caption{We focus on decoding semantic information from fMRI in the auditory neural decoding scenario and use fMRI signals as prompts to guide a pre-trained GPT-2 to achieve decoding.}
    \label{fig: task}
\end{figure}

In recent years, due to the rapid development of deep learning and its widespread application in brain-computer interfaces, especially in neural encoding and decoding \cite{scotti2024reconstructing,lin2022mind,du2023decoding,li2022multi,wang2022multi}, significant progress has been made in the research field of extracting language information from brain signals. For example, reconstructing audio of auditory stimuli from brain signals \cite{yang2015speech,pasley2012reconstructing,santoro2017reconstructing,defossez2023decoding}, or reconstructing corresponding text \cite{affolter2020brain2word,defossez2023decoding,pereira2018toward,xi2023unicorn,tang2023semantic}. In this article, we focus on reconstructing language information by decoding the original text (see Figure \ref{fig: task} for an illustration). Specifically, we focus on text decoding in auditory neural decoding scenarios. For the types of brain-computer interfaces, we focus on decoding text from functional magnetic resonance imaging (fMRI), a widely used non-invasive brain-computer interface in both research and practical applications.

\begin{figure*}[t]
    \centering
    \includegraphics[width=\linewidth]{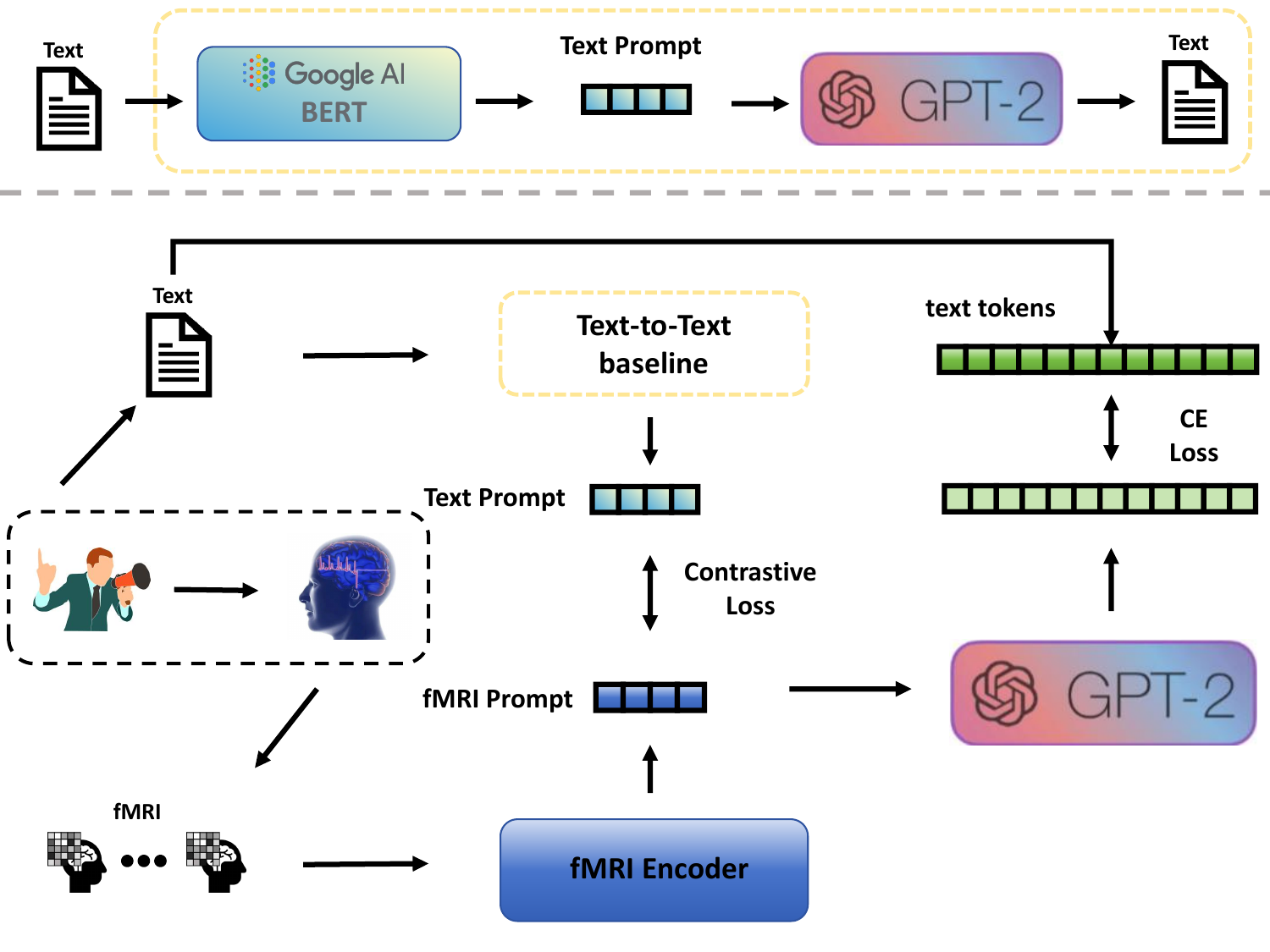}
    \caption{The training stages of our method. The upper part: we use the BERT and GPT-2 for the encoder and decoder of our text-to-text baseline. In this baseline, the BERT representation will be mapped into a text prompt which is used for reconstructing the original text using GPT-2. The lower part: we use a transformer fMRI encoder to extract the fMRI prompt and add a contrastive loss to align the fMRI prompt to the text prompt. Then, the GPT-2 will decode the text according to the fMRI prompt.}
    \label{fig: training}
\end{figure*}

To accomplish this goal, two primary challenges must be addressed. Firstly, the temporal resolution of fMRI signals presents a significant obstacle. Despite its commendable spatial resolution and non-invasive characteristics, fMRI signals exhibit significantly low temporal resolution. For instance, in the context of commonly spoken English, where the average speaking speed exceeds 2 words per second \cite{tang2023semantic}, the BOLD response to neural activity rises and falls over approximately 10 seconds which is extremely slower than the speech stimuli \cite{logothetis2003underpinnings}. Secondly, another essential challenge is the significant difference between fMRI modality and text modality. In fact, in auditory information decoding scenarios, the text does not present as the stimulus signal received by the subject. Rather, it contains the semantics of the stimulus signal. Therefore, the quality of modal alignment significantly influences the effectiveness of the decoding model.

The low temporal resolution in fMRI necessitates our model to decode multiple words from a single fMRI signal during the decoding process. To tackle this ill-posed inverse problem, we propose a solution by incorporating a language prior through a pre-trained Large Language Model (LLM) - Generative Pre-trained Transformer 2 (GPT-2) \cite{radford2019language}. As illustrated in the bottom part of Figure \ref{fig: training}, we establish a mapping from fMRI to prompts and train GPT-2 to employ autoregressive methods, generating corresponding text based on the provided fMRI prompts. By applying the cross-entropy loss to the output logit of GPT-2, the fMRI encoder can learn the suitable prompts for the target text. 

To mitigate the substantial modal disparity between the fMRI signal and the text data, we introduced another text-to-text baseline during the training stage. As depicted in the upper section of Figure \ref{fig: training}, we utilized Bidirectional Encoder Representations from Transformers (BERT) \cite{devlin2018bert} as the text encoder to extract the text prompt. Subsequently, we reconstructed the text using GPT-2 in a manner similar to the brain-to-text. Since the input and output of the text-to-text baseline are identical, eliminating modal differences, we designate the text prompt as the optimal prompt for the ground-truth text. Consequently, we incorporated a contrastive loss to align the fMRI prompt with the text prompt, enhancing decoding performance.

In summary, our main contributions are as follows:
\begin{itemize}
\item We propose \textbf{Brain Prompt-GPT (BP-GPT)}, which is a novel structure that can use the fMRI prompt to decode the text of speech stimuli in an end-to-end structure.

\item We introduced a language prior by a pre-trained LLM (GPT-2) to compensate for the low temporal resolution of fMRI; through contrastive learning, we encourage fMRI prompts to align with text prompts, thereby reducing the impact of modal differences on decoding performance.

\item We evaluated our BP-GPT model on an open-source auditory semantic decoding dataset and achieved a significant improvement of up to $4.61\%$ on METEOR and $2.43\%$ on BERTScore across all subjects compared to the state-of-the-art method. These results demonstrate the feasibility and advantages of our approach.
\end{itemize}

\section{Related Work}
\subsection{Large language models}
Since the introduction of the transformer architecture \cite{vaswani2017attention}, numerous transformer-based Large Language Models (LLMs) have emerged. While all these models utilize stacked attention layers to construct their networks, each exhibits unique characteristics. Broadly speaking, existing LLMs can be categorized into three groups: encoder-only, decoder-only, and encoder-decoder.

The encoder-only LLMs are mostly used to extract features of input text and are used for different downstream tasks. The most representative one is the BERT \cite{devlin2018bert} model. BERT adopts a bidirectional architecture and trains using a masked language model and next-sense prediction task, allowing it to extract representations of text based on the previous and following text. A popular optimized version of BERT is RoBERTa \cite{liu2019roberta}, which builds upon BERT's architecture and pre-training objectives, refining the training process to achieve improved performance. It removes the next sentence prediction objective and trains on more data and for longer epochs, leading to better representations.

Decoder-only LLMs stem from the decoder component of the transformer architecture. Their key feature lies in masking future positions to ensure that predictions of the current token are based solely on preceding tokens. The most prominent example of decoder-only LLMs is the GPT (Generative Pre-trained Transformer) series \cite{radford2018improving,radford2019language,brown2020language,achiam2023gpt}, which introduces a large-scale autoregressive language model based on the Transformer architecture. These models have demonstrated the effectiveness of unsupervised pre-training followed by fine-tuning across various natural language processing tasks. Beyond their technical aspects, they have significantly advanced natural language understanding and generation tasks, including translation, question answering, code generation, and even creative writing. Notably, with the emergence of ChatGPT, the GPT series has become the most popular choice among LLMs.

In addition to the aforementioned architectures, there are models that utilize the encoder-decoder architecture to leverage the strengths of both components. Examples include BART \cite{lewis2019bart} and T5 \cite{raffel2020exploring}. While both models are based on the encoder-decoder architecture, each adopts a unique approach and possesses distinct capabilities for natural language processing and generation tasks. T5 emphasizes a text-to-text framework for unified processing of various tasks, whereas BART focuses on bidirectional and auto-regressive training for sequence-to-sequence tasks such as summarization and translation.

\subsection{Decoding the Brain Signals into Text}
In the early stages of text decoding, model outputs often comprised a single vocabulary and only supported a limited vocabulary set. For instance, in the Brain2word \cite{affolter2020brain2word}, researchers employed a classification model to decode participants' brain signals when they read words. In the study by Defossez et al. \cite{defossez2023decoding}, contrastive learning was applied to decode words or phrases from auditory brain signals. Additionally, Pereira et al. \cite{pereira2018toward} and Sun et al. \cite{sun2019towards} explored the decoding of sentences from brain signals in their respective works.

In recent years, due to the success of LLMs, some work has begun to attempt to use LLMs to decode complete texts from various types of brain data. For example, Wang et al. \cite{wang2022open} proposed using BART \cite{lewis2019bart} to decode text from the EEG and eye tracking signals of the subjects during reading, achieving the decoding of open vocabularies. In DeWave \cite{duan2023dewave}, Duan et al. proposed using a quantized variant encoder and BART to improve the decoding of EEG2text, achieving decoding without external event markers like handling or eye tracking. Unlike decoding using EEG, in UniCoRN \cite{xi2023unicorn}, Xi et al. also used BART to decode fMRI signals into text. Tang et al \cite{tang2023semantic}. used linear regression and GPT, based on the neural encoding architecture, to complete text decoding using similarity measurement. Our work, along with UniCoRN and Tang et al.'s most relevant, is based on fMRI decoding. Different from UniCoRN which treats fMRI as a foreign language and uses machine translation structure for text decoding, we adopted a prompt approach. Compared to Tang et al., which adopts a neural encoding architecture, our method adopts a more direct neural decoding architecture, which is an end-to-end approach.

\begin{figure*}[t]
    \centering
    \includegraphics[width=\linewidth]{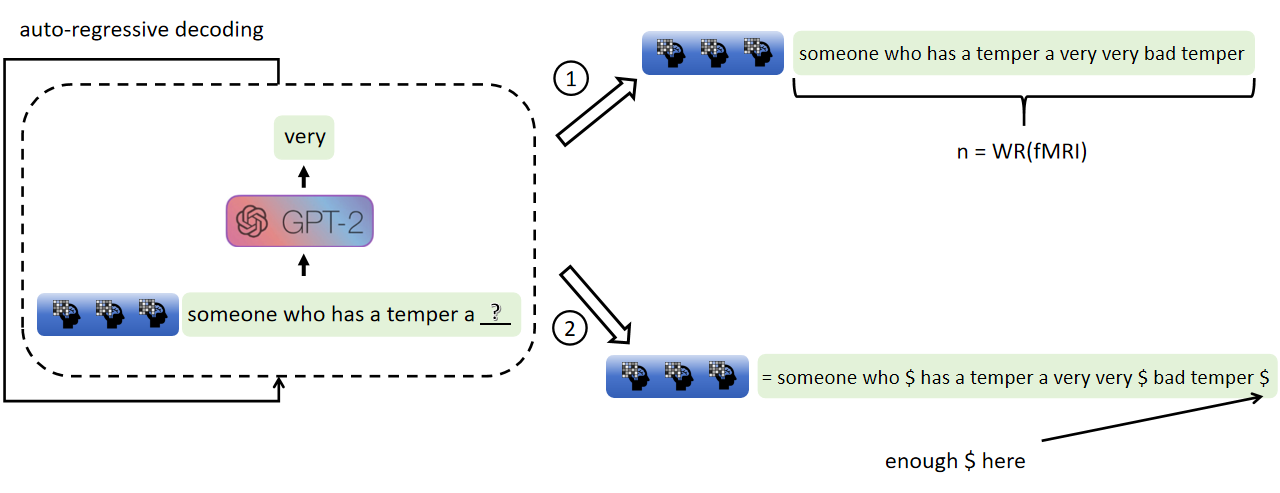}
    \caption{An illustration of the inference stage is provided here. During this stage, the fMRI prompt is considered as the preceding text for the target text generation. Subsequently, GPT-2 generates the text in an autoregressive manner, relying on both the fMRI prompt and the generated text. For deciding the length of decoding text, we compared two strategies in this work. The first one is to use the word rate model to predict the length of text; The second one is to use special tokens and fine-tune the GPT-2, the decoding process will end when GPT-2 generates enough special tokens (\$ in our implementation).}
    \label{fig: inference}
\end{figure*}

\section{Method}
\subsection{A Text to Text Baseline}
The fundamental concept of our method draws inspiration from the encoding and decoding processes applied to text and its corresponding modalities. In the field of Natural Language Processing (NLP) or Computer Vision (CV), various tasks demand the encoding and decoding of text, such as text-to-image generation \cite{ramesh2021zero} image captioning \cite{zhou2020unified,mokady2021clipcap}. In this part, we align both the encoding input and decoding target with the text corresponding to auditory stimuli. Due to the remarkable progress made by LLMs recently \cite{devlin2018bert,radford2019language,brown2020language} and its continued evolution at an astonishing speed  \cite{achiam2023gpt}, we have established the encoder and decoder of our text-to-text model entirely based on the pre-trained LLMs, minimizing the volume of parameters that need to be trained. This design facilitates the seamless integration of new LLMs, enabling easy upgrades for our method.

We chose BERT as the text encoder and GPT-2 as the text decoder in this work and drew inspiration from the recent image caption works \cite{zhou2020unified,mokady2021clipcap} which use encoded representation as the prompt of target text in the decoding. Specifically, we first encode the text into the representation space using the last hidden state of BERT. Then a mapping network is applied to map the BERT representation into the prompt:

\begin{equation}
    \mathit{P}_{i}^{T} = \mathbf{M}_{\theta}(\mathbf{BERT}(x_{i})),
\end{equation}
where the $\mathbf{M}(\cdot)$ denote the mapping network, and $\mathit{P}_{i}^{T}=(p^{T}_{1}, \ldots, p^{T}_{k})$ is the extracted prompt sequence.

To reconstruct the text, we feed the prompt into GPT-2 to generate the original text. The text was generated using the next-token-prediction paradigm and the prompt is considered as the preceding text of the target text. In the training, the parameters of BERT are fixed. For the GPT-2, fine-tuning is not a mandatory option here due to the presence of the mapping network. So only the parameters of the mapping network have to be optimized. We use a cross-entropy loss between the output of GPT-2 and the tokens of the target text for the optimization of the mapping network:

\begin{align}
    \mathcal{L}_{text} &= -\sum_{i=1}^{N}\log p_{\theta}(\mathit{W}|\mathit{P}_{i}^{T})\\
    &= -\sum_{i=1}^{N}\sum_{j=1}^{\mathcal{L}}\log p_{\theta}(w_{j}|p^{T}_{1}, \ldots, p^{T}_{k}, w_{1}, \ldots, w_{j-1}),
\end{align}
where the $\mathit{W} = (w_{1}, \ldots, w_{j})$ is the tokens of the text, and $k$ is the length of prompt.

\subsection{fMRI to Text Decoding}
There are two main challenges in decoding text from the fMRI signals: (1) the low temporal resolution of fMRI and (2) the significant modal difference between fMRI and text. In this part, we will introduce our method which could address the two challenges above. Our method consists of two essential components. The first is the fMRI-prompted text decoding method, which introduces the prompt paradigm into fMRI decoding and addresses the problem of low temporal resolution of fMRI. The second is aligning the fMRI prompt to the text prompt and therefore reduce the impact of modal differences and address the last challenge. 

\subsubsection{fMRI-prompted text decoding.}
Due to the low temporal resolution of fMRI, we need to decode multiple words from each fMRI sample point. We adopted a similar structure as the text-to-text baseline to solve this problem. Specifically, we use a fMRI encoder model to encode the fMRI into the representation space and use this representation as the prompt of the text generation process of GPT-2:

\begin{equation}
    \mathit{P}_{i}^B = \mathbf{E}_{\eta}(x_{i}^B),
\end{equation}
where the $\mathbf{E}_{\eta}$ is the fMRI encoder, and $x_{i}^B$ denote the fMRI. $\mathit{P}_{i}^B = (p^B_{1}, \cdots, p^B_{k})$ denote the fMRI prompt which is extracted by the fMRI encoder.

Since we did not introduce a pre-trained fMRI encoder here, the fMRI-to-text part did not require a mapping network. Furthermore, the fMRI encoder can be trained using the cross-entropy loss on the output of the GPT-2. Here, fine-tuning the GPT-2 is not a mandatory option either. When we choose not to fine-tune the GPT-2, the loss is completely dependent on the parameters of the fMRI encoder $\eta$ and has the following form:
\begin{align}
    \mathcal{L}_{brain} &= -\sum_{i=1}^{N}\log p_{\eta}(\mathit{W}|\mathit{P}_{i}^B)\\
    &= -\sum_{i=1}^{N}\sum_{j=1}^{\mathcal{L}}\log p_{\eta}(w_{j}|p^B_{1}, \ldots, p^B_{k}, w_{1}, \ldots, w_{j-1}).
\end{align}

However, fine-tuning the GPT-2 may bring performance improvements in downstream tasks. Meanwhile, in the text decoding task of this work, the fine-tuning of the GPT-2 can bring more specific benefits. We will introduce this specific benefits in Section \ref{sec: inference} and compare these two options in the later experimental section.

\subsubsection{Align with the Optimal Prompt}
Given the considerable significant modal differences between fMRI and text, extracting effective fMRI prompts through the fMRI encoder poses challenges. Conversely, the text baseline inherently circumvents this modal difference. So we argue that the prompt derived from the text baseline can be deemed as the optimal prompt for the text to be generated. To utilize the optimal prompt, we employ contrastive learning to align the fMRI prompt with the text prompt.

Specifically, we set the fMRI prompt and text prompt of the same text as the positive pair and calculate the similarity between the positive pair using the following formula:

\begin{equation}
    \mathit{S}_{p} = \exp (cos(\mathit{P}^{i}_B \cdot \mathit{P}^{i}_T)/\tau),
\end{equation}
where the $\tau$ refers to the temperature hyperparameter.

For the negative pair, we use the fMRI prompt and text prompt from different texts and the fMRI prompt of different texts. The similarity between the negative pairs can be formulated as:

\begin{equation}
\begin{split}
    \mathit{S}_{n} = &\exp (cos(\mathit{P}^{i}_B \cdot \mathit{P}^{j}_B)/\tau) +  \exp (cos(\mathit{P}^{i}_B \cdot \mathit{P}^{j}_T)/\tau), i\neq j.
\end{split}
\end{equation}

Based on the definition above, the contrastive loss has the following form:

\begin{equation}
    L_{\mathcal{C}} = -\mathbb{E}\left[\log\frac{S_p}{S_n}\right].
\end{equation}

\subsection{Training}
We will divide the training process into two stages. In the initial stage, the text-to-text baseline is trained for text encoding and decoding. This process enables the model to learn to extract the optimal prompt for the target text, which is then utilized as the target of contrastive learning in the subsequent stage. Then, we train our decoding model in the second stage using the following formula:
\begin{align}
    L = L_{brain} + \alpha L_{C},
\end{align}
where the $\alpha$ is a hyperparameter for the contrastive loss.

\subsection{Inference}\label{sec: inference}
As is shown in Figure \ref{fig: inference}, in the inference stage, we extract the fMRI prompt first and generate the text using GPT-2 depending on the fMRI prompt. The text will be generated word by word in a next-token-prediction paradigm. 

As our focus is on text decoding within an auditory neural decoding scenario, the auditory stimuli received by the subjects exclusively consist of words without punctuation. This will cause trouble for the model during the inference phase, as we cannot determine the end of generation through the stop token (usually, periods are used). Although punctuation can be manually added during text annotation for audio stimuli, this introduces two concerns. Firstly, the speaker might have delivered a spontaneous speech without adhering to the ground-truth speech draft. Secondly, since the decoding often uses the fMRI signals within a fixed-length window, and the end of this window does not always match the end of the sentence, even if we add the punctuation manually, they are very likely unable to indicate the end of the generation.

For the reasons outlined above, we chose not to include punctuation in the model training process. To find the end of the generation in the inference stage, we offer two strategies here, an illustration of these two strategies is shown in Figure \ref{fig: inference}.

In the first one, we adopt an approach from recent work \cite{tang2023semantic}, utilizing a word rate model to predict the number of words perceived by participants. The text generation process will be stopped when the length of the generated text meets the word count predicted by the word rate model.

The second strategy is to use the special token to segment text based on the repetition time (TR) of fMRI. In this work, we add \$ to the gound-truth text during the training, and stop the generation when we meet enough \$ in the inference text. We also add an equal mark to point out the beginning of the gourd-truth text.

\subsection{Mapping Network and fMRI Encoder Model}
The core of our model lies in mapping the original fMRI signal or representation of the text encoder into the prompt. Given the complexities in these transformations, especially for the fMRI-to-text part, we employed the transformer structures \cite{vaswani2017attention} to accomplish the mapping of original features to the prompt. The transformer's inherent capability for global attention enables the extraction of semantic relationships from the input fMRI sequence, yielding a fMRI prompt better suited for the target text.

\begin{table*}[t]
\resizebox{\linewidth}{!}{\begin{tabular}{c|ccc|ccc|ccc}
\hline
\multirow{2}{*}{}  & \multicolumn{3}{c}{\textbf{BLEU-1}$\uparrow$} & \multicolumn{3}{c}{\textbf{METEOR}$\uparrow$} & \multicolumn{3}{c}{\textbf{BERTScore}$\uparrow$} \\ \cline{2-10} 
                  & UTS01       & UTS02       & UTS03      & UTS01       & UTS02       & UTS03      & UTS01        & UTS02       & UTS03       \\ \hline
T2T+WR               & 0.1968  &   0.2085    &  0.1862   &  0.1414   &   0.1466    &   0.1266
  &   0.8192  &   0.8205  &   0.8163
  \\ \hline
T2T+WR+fine-tune               & 0.2296  &   0.2421    &  0.241   &  0.1692   &   0.1699    &   0.176
  &   0.8242  &   0.8278  &   0.8259
  \\ \hline
T2T+Spe               & 0.2189  &  0.2044    &  0.2187   &  0.2032   &   0.204    &   0.2082
  &   0.8325  &   0.8328  &   0.8343
  \\ \hline  
T2T+Spe+fine-tune               & \textbf{0.2621}  &   \textbf{0.2613}    &  \textbf{0.2554}   &  \textbf{0.2627}   &   \textbf{0.2597}    &   \textbf{0.2549}
  &   \textbf{0.8432}  &   \textbf{0.8451}  &   \textbf{0.8417}
  \\ \hline
\end{tabular}}
\caption{Text-to-text performance. 'WR' refers to use word rate model in the inference, and 'Spe' refers to the special tokens. In the results without annotation 'fine-tune', the parameter of GPT-2 is fixed.}
\label{tab: T2T-performance}
\end{table*}

\begin{figure}[t] 
    \centering
    \subfigure[UTS01]{%
        \includegraphics[width=0.45\textwidth]{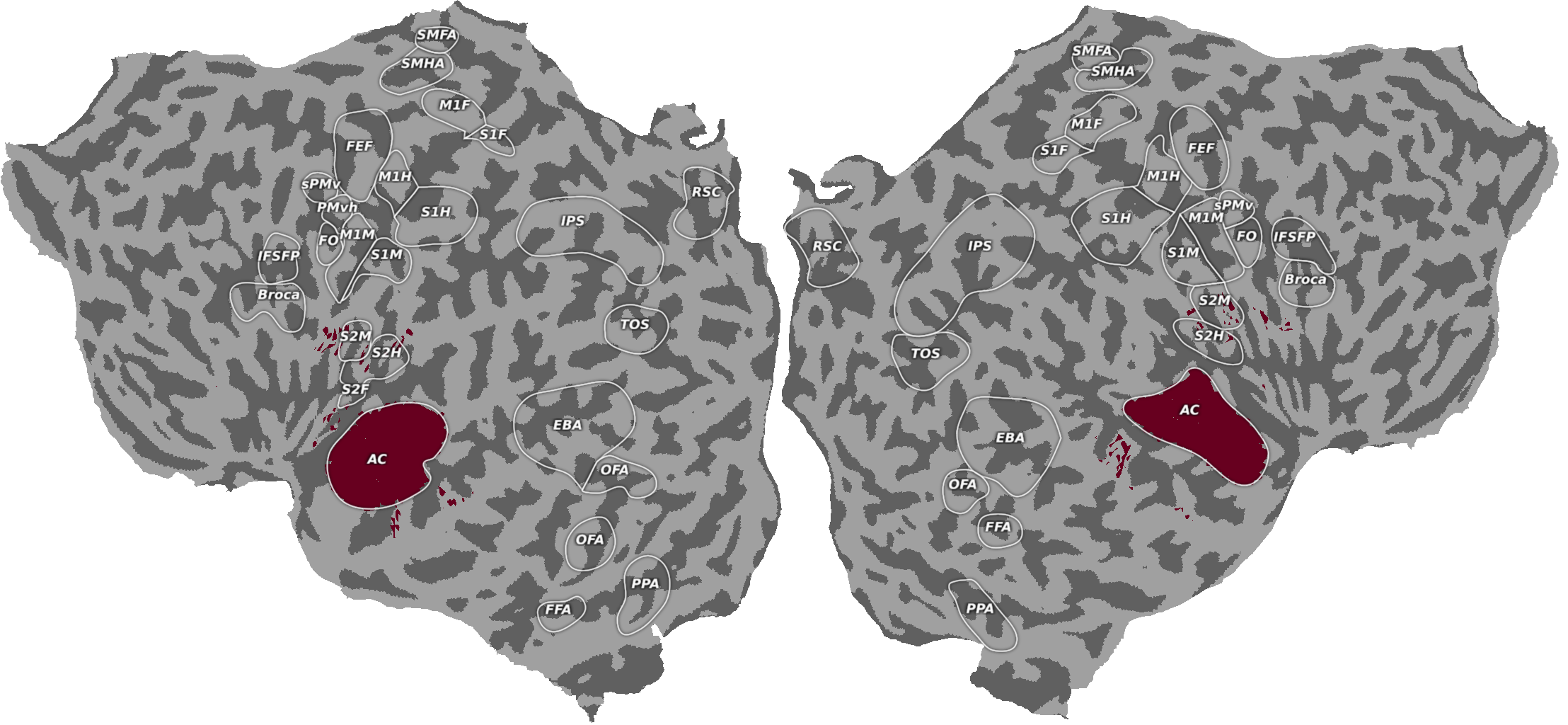}%
    }
    \quad
    \subfigure[UTS02]{%
        \includegraphics[width=0.45\textwidth]{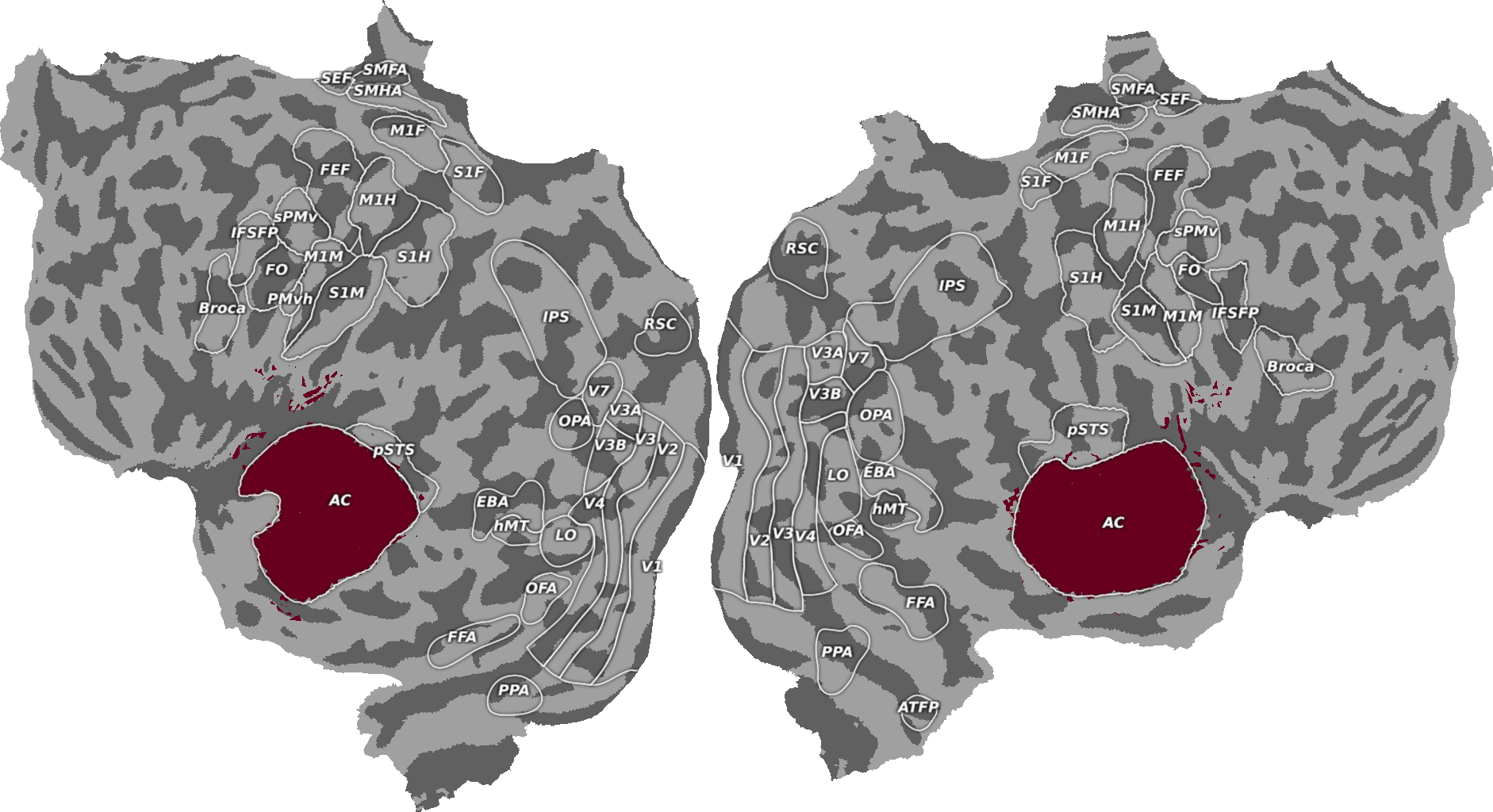}%
    }
    \quad
    \subfigure[UTS03]{%
        \includegraphics[width=0.45\textwidth]{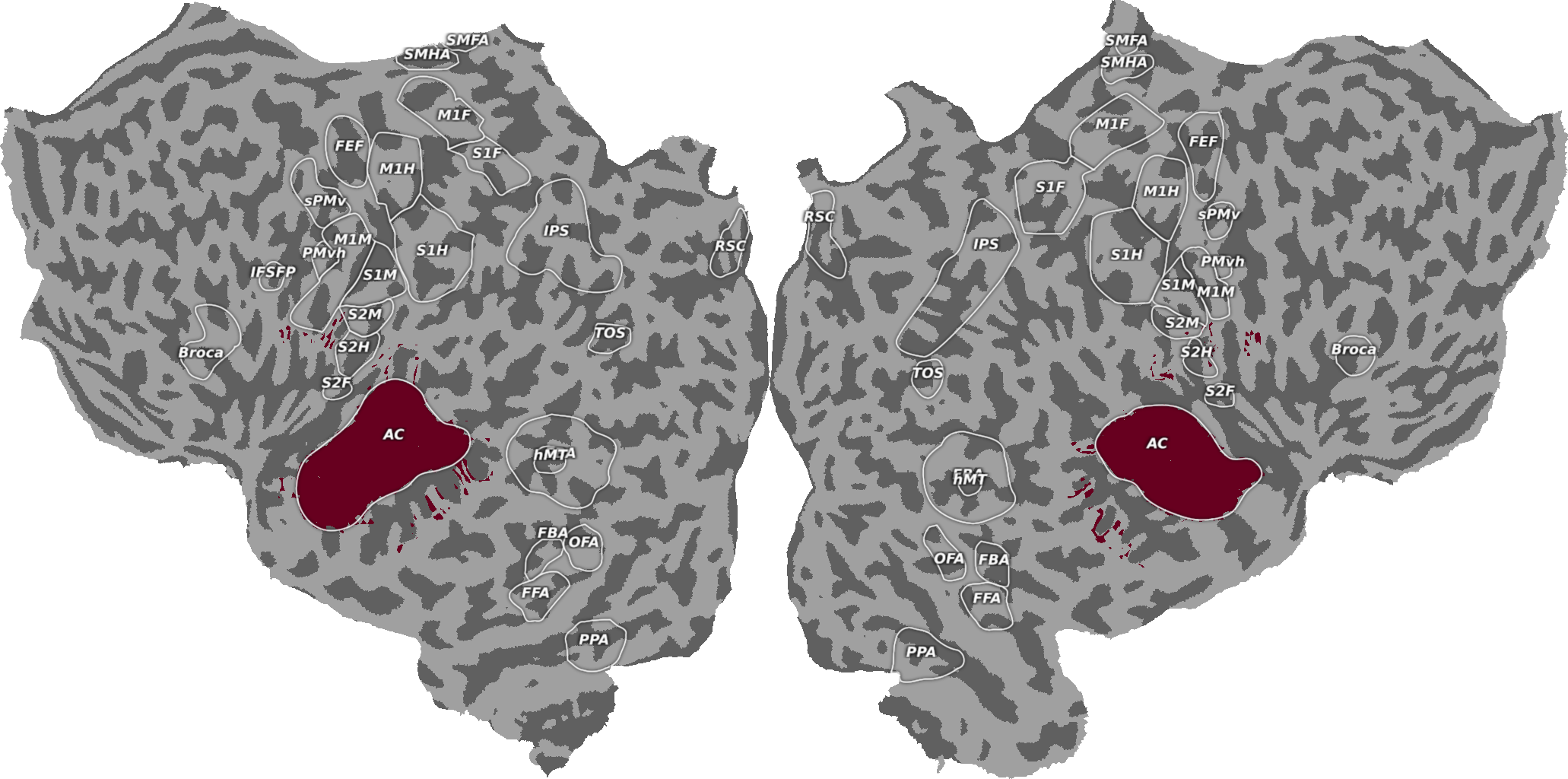}%
    }
    \caption{The cortical flatmaps for the auditory cortex (in red color) of the different subjects we used.}\label{fig: ROI} 
\end{figure}

\section{Experiment}
\subsection{Dataset}
We evaluate our method on an fMRI dataset obtained during a passive natural language listening task \cite{lebel2023natural} along with its extended data \cite{tang2023semantic}. This dataset comprises fMRI data from 8 subjects recorded while they passively listened to naturally spoken English stories. The stories were sourced from \textit{The Month} and \textit{New York Times Modern Love} podcasts. Specifically, the first 3 subjects (UTS01 to UTS03) in the dataset had access to stories from both \textit{The Month} and \textit{New York Times Modern Love}, expanding the total number of stories to 84 for these subjects. For the remaining 5 subjects (UTS04 to UTS08), all 27 stories are from \textit{The Month}. To maintain consistency with existing works \cite{lebel2023natural,tang2023semantic}, we selected the story "Where There’s Smoke", shared by all subjects, as the test set.

In the vanilla section of the dataset, the stories were divided into 5 sections, each containing 5 stories and an additional test story ("Where There’s Smoke"). This configuration resulted in a total duration of stimulating audio exceeding 6 hours for all subjects. In the extended section of the dataset, the first three subjects underwent an additional 10 sessions, increasing the overall dataset duration to 81 hours across all subjects. All stories feature a single speaker narrating an autobiographical tale without a prepared script. The texts were manually transcribed by one listener and automatically aligned to the audio using the Penn Phonetics Lab Forced Aligner (P2FA) \cite{yuan2008speaker}. Consequently, the transcriptions lack punctuation marks, and some words may be repeated as the speaker is thinking and organizing their language. Additionally, certain sounds, such as "cough," "laugh," "lip smack," "misc noise," and "silence," were annotated.

\subsection{Implementing Details}
In our work, we choose the first three subjects that have the extended data for all of our experiments. For the region of interest (ROI), we use the voxel in the auditory cortex for our experiment (see Figure \ref{fig: ROI} for the cortical flat maps). The temperature of contrastive loss is set to $\tau=0.1$, and the weight of contrastive loss is set to $\alpha=1$. We split the fMRI sequence and corresponding text into multiple 20-second windows, with zero overlaps. For the prompt length, we use $k=30$.

For the mapping network, the BERT representation will be first mapped to a 512-dimensional vector before passing forward to the transformer. We use an 8-layer transformer here, with 8 attention heads in each layer. The fMRI encoder has the same architecture as the mapping network, except for a linear layer for the input.

All the codes are implemented using PyTorch \cite{paszke2019pytorch}, and training on an Nvidia A-100 GPU with AdamW optimizer \cite{loshchilov2017decoupled}. The batch size is set to 32.

\subsection{Baseline and Evaluation Metrics}
We compare our method to Tang et al. \cite{tang2023semantic}, which is the state-of-the-art method in the dataset we used.  In their method, they use a neuron encoding structure, which uses the GPT to generate proposal words and encode the proposal words to fMRI to find the most matching word. For a fair comparison, we use the same story ("Where There’s Smoke") in the dataset as the test set and divide the entire story into 20-second windows, calculate evaluation metrics within each window, and take the average of all windows as the metric score for the entire test set. 

Similar to Tang et al. \cite{tang2023semantic}, we use identical language similarity metrics to evaluate our method in several aspects. BLEU \cite{papineni2002bleu} indicates the number of individual translated segments that appear in the ground-truth text. METEOR \cite{denkowski2014meteor} computes the harmonic mean of unigram precision and recall. And BERTScore \cite{zhang2019bertscore} computes a similarity score for each token in the candidate sentence with each token in the reference sentence using contextual embeddings.

\begin{table*}[t]
\resizebox{\linewidth}{!}{\begin{tabular}{c|ccc|ccc|ccc}
\hline
\multirow{2}{*}{}  & \multicolumn{3}{c}{\textbf{BLEU-1}$\uparrow$} & \multicolumn{3}{c}{\textbf{METEOR}$\uparrow$} & \multicolumn{3}{c}{\textbf{BERTScore}$\uparrow$} \\ \cline{2-10} 
                 & UTS01       & UTS02       & UTS03      & UTS01       & UTS02       & UTS03      & UTS01        & UTS02       & UTS03       \\ \hline
Tang et al.        & \textbf{0.2331}     & \textbf{0.2426}     & \textbf{0.2470}    & 0.1621     & 0.1677     & 0.1703    & 0.8077      & 0.8104      & 0.8116     \\ \hline
BP-GPT               & 0.2159  &   0.2111    &  0.2113   &  \textbf{0.2082}   &   \textbf{0.1976}    &   \textbf{0.2034}
  &   \textbf{0.832}  &   \textbf{0.8322}  &   \textbf{0.8332}
  \\ \hline
\end{tabular}}
\caption{Compare our method with the existing work. The \textbf{BP-GPT} refers to using the special tokens for the inference and fine-tuning the GPT-2.}
\label{tab: performance}
\end{table*}

\subsection{Evaluation the Text-to-text Baseline}\label{sec: exp-T2T}
Since we treat the text prompt as the optimal prompt for decoding in our method, we first evaluate the performance of the text-to-text baseline in our work. We consider 4 settings in this part. For the first two settings, we use the word rate model in the inference and fine-tune or fix the parameters of GPT-2 in the training. For the last two settings, we use the special tokens and also compare two options for fine-tuning or fixing the parameters of GPT-2.

The experiment results are listed in Table \ref{tab: T2T-performance}. From the experimental results, we can find that our text-to-text baseline can effectively encode and decode text under various experimental settings. This result supports our further application of this prompt paradigm in fMRI-to-text decoding. Also, we find that fine-tuning the GPT-2 in the training can always bring improvements in performance, regardless of the inference strategy we used. For the inference strategies, rather than using a word rate model to infer the text length, we find that adding special tokens in the ground-truth text in the training stage can effectively improve the decoding performance. Moreover, we find that fine-tuning the GPT-2 for the special tokens can bring a significant improvement in the results. Specifically, this setting can bring up to $5.69\%$ on BLEU-1, $5.95\%$ on METEOR, and $1.23\%$ on BERTScore among all the subjects. Based on these experiment results, we choose the setting that uses special tokens and fine-tune the GPT-2 as the text-to-text baseline for all the following experiments which use the special tokens as inference strategies. Also, the text-to-text baseline for the word rate model strategies in the following experiments is also fine-tuned.

\subsection{Evaluation of fMRI to Text Decoding}
In this part, we evaluate the decoding performance of our model. To align with existing work \cite{tang2023semantic}, we report the performance of the first three subjects: UTS01, UTS02, and UTS03, who have experienced all 84 stories in the experiment. For the setting of our method, we add special tokens in the ground-truth text for the training. The GPT-2 is fine-tuned in the training, and the inference is stopped when the GPT-2 generates enough \$. We refer to this setting as the \textbf{BP-GPT} in all the experiments.

As is shown in Table \ref{tab: performance}, our method achieves comparable or even better performance. Specifically, on the METEOR, our method can achieve an improvement ranging from $2.99\%$ to $4.61\%$ on all the subjects. For the BERTScore, we also achieve an improvement ranging from $1.87\%$ to $2.43\%$ on all the subjects. 

\subsection{Ablation Study}
In this part, we make ablation studies that evaluate the contributions of contrastive learning and different inference strategies for the performance. We report these experiment results in Table \ref{tab: Abalation-contrastive} and Table \ref{tab: Abalation-inference}. For the inference strategies, we mark 'WR' for the results that use the word rate model to distinguish it from the results that use the special tokens in the inference.

\begin{table*}[t]
\resizebox{\linewidth}{!}{\begin{tabular}{c|ccc|ccc|ccc}
\hline
\multirow{2}{*}{}  & \multicolumn{3}{c}{\textbf{BLEU-1}$\uparrow$} & \multicolumn{3}{c}{\textbf{METEOR}$\uparrow$} & \multicolumn{3}{c}{\textbf{BERTScore}$\uparrow$} \\ \cline{2-10} 
                  & UTS01       & UTS02       & UTS03      & UTS01       & UTS02       & UTS03      & UTS01        & UTS02       & UTS03       \\ \hline
BP-GPT-wo-contras (WR)         & 0.1855  &   0.1851    &  0.1979   &  0.1372  &   0.1358    &   0.1442  &   0.8134  &   0.8131 &   0.8172
  \\ \hline               
BP-GPT (WR)         & 0.2052  &   0.1944    &  0.2017   &  0.1451  &   0.1476    &   0.1497  &   0.8192  &   0.8164  &   0.8198
  \\ \hline
BP-GPT-wo-contras         & 0.2027  &   0.2041    &  0.2043   &  0.1943  &   0.1958   &   0.1946  &   0.829  &   0.8278  &   0.8281
  \\ \hline
BP-GPT               & \textbf{0.2159}  &   \textbf{0.2111}    &  \textbf{0.2113}   &  \textbf{0.2082}   &   \textbf{0.1976}    &   \textbf{0.2034}
  &   \textbf{0.832}  &   \textbf{0.8322}  &   \textbf{0.8332}
  \\ \hline
\end{tabular}}
\caption{Abalation study: contrastive learning. Experiments that use the word rate model in the inference are marked with 'WR'.}
\label{tab: Abalation-contrastive}
\end{table*}

\begin{table*}[t]
\resizebox{\linewidth}{!}{\begin{tabular}{c|ccc|ccc|ccc}
\hline
\multirow{2}{*}{}  & \multicolumn{3}{c}{\textbf{BLEU-1}$\uparrow$} & \multicolumn{3}{c}{\textbf{METEOR}$\uparrow$} & \multicolumn{3}{c}{\textbf{BERTScore}$\uparrow$} \\ \cline{2-10} 
                  & UTS01       & UTS02       & UTS03      & UTS01       & UTS02       & UTS03      & UTS01        & UTS02       & UTS03       \\ \hline
BP-GPT-wo-fine-tune (WR)         & 0.198  &   0.1936    &  0.1997   &  0.1343  &   0.1409    &   0.139  &   0.8157  &   0.8145  &   0.8129
  \\ \hline
BP-GPT (WR)         & 0.2052  &   0.1944    &  0.2017   &  0.1451  &   0.1476   &   0.1497  &   0.8192  &   0.8164  &   0.8198
  \\ \hline
BP-GPT-wo-fine-tune         & 0.2083  &   0.1949    &  0.2065   &  0.206  &   0.1924    &   0.2022  &   0.8318  &   0.8295  &   0.8298
  \\ \hline
BP-GPT               & \textbf{0.2159}  &   \textbf{0.2111}    &  \textbf{0.2113}   &  \textbf{0.2082}   &   \textbf{0.1976}    &   \textbf{0.2034}
  &   \textbf{0.832}  &   \textbf{0.8322}  &   \textbf{0.8332}
  \\ \hline
\end{tabular}}
\caption{Abalation study: inference strategy. Experiments that use the word rate model in the inference are marked with 'WR'.}
\label{tab: Abalation-inference}
\end{table*}

\begin{figure}[b] 
    \centering
    \subfigure[BLEU-1 and METEOR]{%
        \includegraphics[width=0.45\textwidth]{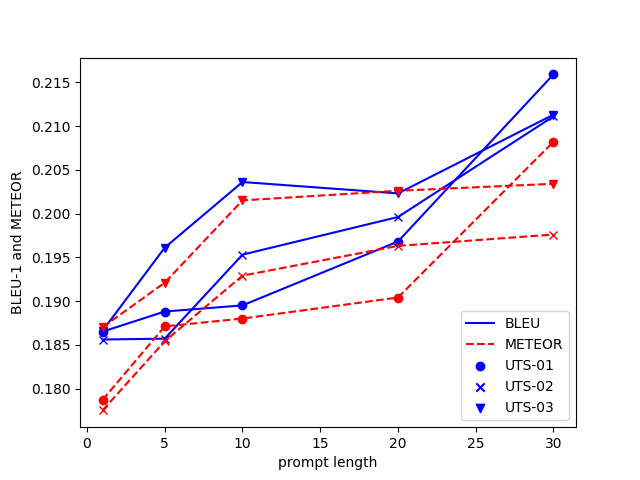}%
    }
    \subfigure[BERTScore]{%
        \includegraphics[width=0.45\textwidth]{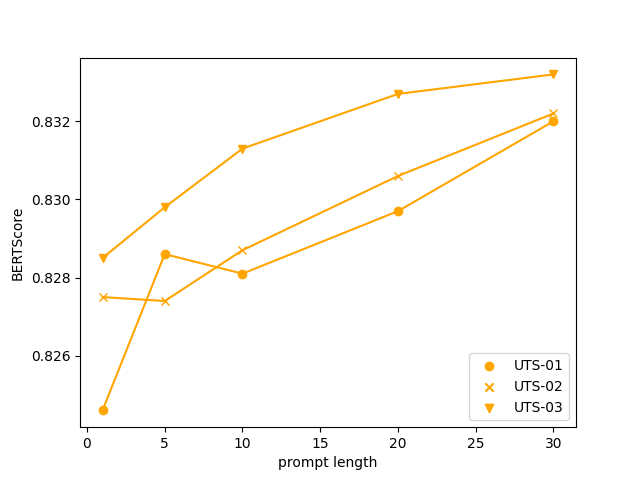}%
    }
    \caption{The performance under different prompt length.}\label{fig: prompt} 
\end{figure}

\subsubsection{Contrastive Learning.}\label{sec: contras} To demonstrate the effectiveness of contrastive learning, we compare the performance of our method with or without contrastive learning. We include the experiment results on both inference strategies since the challenge of the significant modal difference between the fMRI and text exists both in these settings. We want to explore through these experiments whether aligning the fMRI prompt with the text prompt can bring performance improvements in various inference strategies. Here, different inference strategies will lead to different text-to-text baselines. As the target of the conservative learning, the text prompt is extracted using the text-to-text baseline that has the same inference strategy as the fMRI-to-text model. We would like to refer the Section \ref{sec: exp-T2T} for more details.

We report the experiment results in Table \ref{tab: Abalation-contrastive}. By comparing Table \ref{tab: Abalation-contrastive} with Table \ref{tab: T2T-performance}, we can find that there is a big gap between the performance of fMRI-to-text decoding and the performance of the text-to-text decoding, indicating the modal differences between text and fMRI impact the decoding performance seriously. Also, through the results, we find that aligning the fMRI prompt with the text prompt always brings performance improvements, no matter what inference strategies have been chosen. Specifically, when using a word rate model at the inference stage, aligning the fMRI prompt to the text prompt can bring an improvement up to $1.97\%$ on BLEU-1, $1.18\%$ on METEOR, and $0.58\%$ on BERTScore. While choosing the special tokens for inference, contrastive learning can bring an improvement up to $1.32\%$ on BLEU-1, $1.39\%$ on METEOR, and $0.44\%$ on BERTScore. This result proves that our approach of aligning fMRI prompts with text prompts is feasible and effective. 

\subsubsection{Inference Strategy.} Due to the characteristics of decoding tasks in auditory decoding scenarios, the choice inference strategy has become particularly important. We further take the ablation study on it. Specifically, we compared four experimental settings, including the performance of two inference schemes with a fine-tuned and not fine-tuned GPT-2. The result is reported in Table \ref{tab: Abalation-inference}. Also, same as in Section \ref{sec: contras}, the aligning target is the corresponding text prompt under the same inference strategy.

As is shown in Table \ref{tab: Abalation-inference}, we find that using special tokens to indicate the end of decoding can always bring a performance improvement. Also, fine-tuning the GPT-2 can bring more improvement with the special tokens. We believe that fine-tuning the GPT-2 in the training can make the parameters of both the fMRI encoder and GPT-2 adapt to the special token. However, if fine-tuning is not performed, only the fMRI encoder will learn to adjust the fMRI prompt to enable GPT-2 to output \$ at the end of each text fragment that corresponds to a fMRI TR, bringing a negative influence on the performance.

\subsection{Prompt Length}
In our method, fMRI drives GPT-2 to generate decoding targets through the prompt. Therefore, the length of the prompt is crucial for the decoding performance. In this part, we investigated the relationship between prompt length and decoding performance and reported the result in Figure \ref{fig: prompt}. We use our BP-GPT setting, that is use special tokens in inference with a fine-tuned GPT-2 model. 

As is shown in the figure, the decoding performance increases with the length of the prompt. Although a longer prompt length can bring better results, it also incurs greater hardware costs, whether it is for the mapping network or the fMRI encoder. Due to GPU memory limitations, the maximum prompt limit during our experiment was 30. However, based on the experimental results, we expect a better performance with a longer prompt length.

\subsection{Conclution and Future Works}
In this work, we propose a decoding method capable of extracting text from fMRI signals within the auditory neural decoding scenario. The central concept of our method involves employing an fMRI-prompted Large Language Model (LLM) for decoding. Specifically, we utilize an fMRI encoder to extract fMRI representations, which serve as prompts for the pre-trained GPT-2 model. Through the application of cross-entropy loss, our fMRI encoder learns the appropriate prompt for GPT-2 to generate the target text. In order to reduce modal differences between the text and fMRI, we introduce a contrastive loss to align the fMRI prompt with the text prompt. Experimental results demonstrate the effectiveness and advantage of our approach.

In summary, our prompt-based LLM decoding method offers ease of implementation and extends to various text-based neural decoding tasks. Furthermore, with the ongoing advancement of LLMs, our method remains readily compatible with updated and superior LLMs, facilitating performance improvements effortlessly. Moving forward, our focus will be on applying our prompted LLM decoding paradigm to a broader range of neural decoding fields and integrating it with additional LLMs.


\bibliographystyle{ACM-Reference-Format}
\bibliography{sample-base}

\end{document}